# Transition Watchpoints: Teaching Old Debuggers New Tricks

Kapil Arya[a,d], Tyler Denniston[b], Ariel Rabkin[c], and Gene Cooperman[d]

a   Mesosphere, Inc., Boston, MA, USA
b   Massachusetts Institute of Technology, Cambridge, MA, USA
c   Cloudera, Cambridge, MA, USA
d   Northeastern University, Boston, MA, USA

**Abstract**   Reversible debuggers and process replay have been developed at least since 1970. This vision enables one to execute backwards in time under a debugger. Two important problems in practice are that, first, current reversible debuggers are slow when reversing over long time periods, and, second, after building one reversible debugger, it is difficult to transfer that achievement to a new programming environment.

The user observes a bug when arriving at an error. Searching backwards for the corresponding fault may require many reverse steps. Ultimately, the user prefers to write an expression that will transition to false upon arriving at the fault. The solution is an expression-transition watchpoint facility based on top of snapshots and record/replay. Expression-transition watchpoints are implemented as binary search through the timeline of a program execution, while using the snapshots as landmarks within that timeline. This allows for debugging of subtle bugs that appear only after minutes or more of program execution. When a bug occurs within seconds of program startup, repeated debugging sessions suffice. Reversible debugging is preferred for bugs seen only after minutes.

This architecture allows for an efficient and easy-to-write snapshot-based reversible debugger on top of a conventional debugger. The validity of this approach was tested by developing four personalities (for GDB, MATLAB, Perl, and Python), with each personality typically requiring just 100 lines of code.

## The Art, Science, and Engineering of Programming

**Transition Watchpoints**

## 1  Introduction

Popular programming environments generally include a debugging tool, with which developers can inspect the flow of execution and the contents of program variables. Virtually every debugger supports forward execution of programs (via "step" or "next" commands.) Many debuggers offer breakpoints or watchpoints to find program points of interest. A few, known as reversible debuggers, also allow the developer to step backwards in the execution history. But those are tied to a specific programming language or a specific debugger. This paper will describe how reverse functionality, and searches over a program history, can be added to arbitrary existing debuggers *without modification to the debugger or the underlying programming infrastructure*.

While forward debugging is routine, reverse debugging is supported by only a handful of debugging tools. Recent versions of GDB include support for reverse; this relies on instruction-level logging and therefore imposes a very large performance penalty on the initial forward execution. Faster implementations of reversibility rely on deterministic record-replay technologies [6, 11, 12, 20, 25, 27, 28, 29, 32, 37, 38, 45]. Incorporating reversibility into a debugging platform is a major engineering effort.

This paper shows how to implement reversibility in an efficient and "universal" way that can be retargeted to any existing debugger. By using the existing (underlying) debugger as an architectural primitive along with a replay framework, it is possible to add reverse and replay to an existing debugging environment much more easily. We call this approach *composite debugging*.

Simply adding a reverse-step feature to an existing debugger is useful, but not sufficient. When debugging, users are seeking to find the point in their program where the program behavior starts to deviate from their expectation of how the program should behave. Stepping through the program can be tedious or entirely infeasible for complex or long-running software. As a result, it is valuable to have higher-level abstractions for finding crucial points in a program's execution.

Many debuggers offer a watchpoint feature, which stops the program when a particular variable or memory location changes its value. We extend this notion by defining *expression-transition watchpoints*, which stop the program at the point where the value of an expression, not merely a variable, changes. A naive implementation would evaluate the expression at each program step. This is the approach of GDB versions 7.0 and higher [14].

Unfortunately, repeatedly checking an invariant can be prohibitively expensive. Consider a program that maintains a complex data structure such as a circular linked list embedded inside some other structure. Suppose a bug sometimes corrupts this list. A programmer might like to see the first point where the list invariant is violated. In a multithreaded program, this might be formalized as "the first point where the list is not circular and where the relevant lock is not set." This cannot be expressed as a simple memory watchpoint, since the list has many nodes, and those nodes may change dynamically. We use the ability to rewind and replay execution to search efficiently for points where the expression changes value. This allows developers to find the exact point in execution where an expression changes value, without paying an exorbitant runtime cost.

This paper makes three contributions:
- We describe how to build a reversible debugger for any language having a conventional debugger, by utilizing replay. This is discussed in the next section.
- We describe and evaluate expression-transition watchpoints. The design is described in Section 3, and the evaluation is in Section 6.
- We describe and evaluate a concrete implementation of these features: FReD, the **F**ast **Re**versible **D**ebugger. We go into detail about the specific engineering decisions and tradeoffs required to make our architecture work with complex existing software systems, such as GDB, which have a number of quirks. These details are described in Section 4.

This work builds on two earlier workshop contributions [4, 42]. This paper unifies and systematizes some of those ideas, and describes an actual working implementation that supports multiple underlying debuggers. The software is available as open source at http://github.com/fred-dbg/fred.

## 2 Design

To date, debuggers have largely been engineered as black-box systems. One of the contributions of this work is showing how features can be added to existing debuggers, in a black-box way. We refer to this technique as *composite debugging* and believe it is of independent value, beyond enabling replay debugging.

We describe a *composite debugger* architecture, which adds a replay debugger on top of a traditional underlying debugger. The underlying debugger may be a separate process (e.g., GDB) or it may be integrated into the language itself as is typically the case for interpreted languages. (See Figure 1.)

Reversibility is achieved through a combination of checkpoints and deterministic forward execution. A "reverse" is implemented by picking the last checkpoint before the target of the reverse, and then walking the program forward to the target point. Inside the deterministic-replay environment, the underlying debugger is used to control the forward execution of the target program. An important and unobvious consequence of this design is that executing a "reverse" will modify the state of the debugger, not only of the target program. This is desirable, since it means that the debugger state and the program state stay synchronized. It also means that if a user has modified the program state via the debugger, and then calls `reverse`, those modifications will be undone, which matches the intuitive meaning of "reverse."

The underlying debugger is specialized to the language of the program being debugged. (For example, the Python debugger for Python programs, or GDB for C programs.) The replay debugger then controls the underlying debugger in order to step the program forwards. We can decompose "continue" and "next" into repeated "step" as needed to arrive at a particular point in time [42].

The remainder of this section describes more particularly the assumptions we make about the underlying debugger, the requirements we have for the replay layer, and the details of how we use replay to implement reverse.

**Transition Watchpoints**

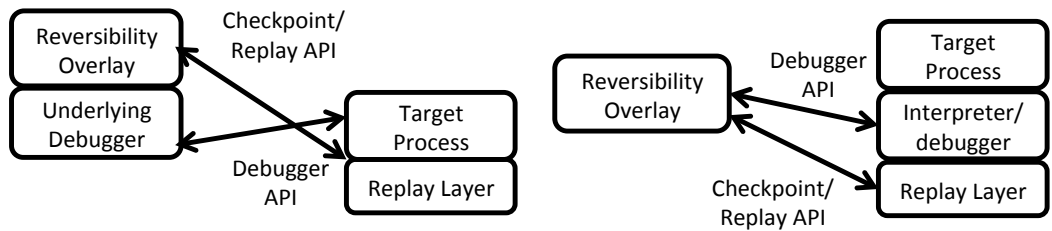

**Figure 1** The composite debugger architecture, for compiled (left) and interpreted (right) processes. Different architectures are required depending on whether the debugger is part of the process being debugged.

## 2.1 The Underlying Debugger

We make use of the underlying debugger in several ways. As discussed above, we use the debugger's step commands to control the state of the target process. The underlying debugger also supplies facilities for inspecting program state. In normal operation, the replay tool is transparent to the end user, and allows the user to interact with the composite debugger in the same way as without the replay tool. The full range of underlying debugger features remains available.

Different debuggers have different user interfaces, capabilities, and methods of interaction with the user and the program being debugged. In order to be widely applicable, the replay layer must be easily adapted to different underlying debuggers. We make only minimal assumptions about the underlying debuggers. In particular we assume step-in (called "step" by GDB), step-over (called "next" by GDB), and breakpoint/continue commands.

**Table 1** Comparison of debugger features. Step commands are ubiquitous; watchpoints and reverse are rare.

| Debugger | step-in, over, and out | conditional breakpoints | Scheduler-locking | value watchpoints | Reverse mode |
|---|---|---|---|---|---|
| GDB | yes | yes | yes | yes | yes |
| lldb | yes | yes | no | yes and conditions | no |
| pydb | yes | yes | no | no | no |
| jdb | yes | no | no | no | no |
| Matlab | yes | yes | no | no | no |
| Eclipse | yes | yes | no | on fields, not specific objects | no |

Typically, when performing step-in/over in a multi-threaded program, the background (non-active) threads are allowed to run freely until the foreground (active) finishes executing the intended instructions (or hits a breakpoint). The background threads could change the state of some shared data, or if they hit a breakpoint, could steal focus from the foreground thread. GDB provides *scheduler-locking* for locking the OS scheduler to allow only a single thread to run. Without scheduler-locking, one cannot isolate the effects of executing a single instruction by a given thread.

Different debuggers use different command syntax, and so the reversible debugger needs a translation layer that is customized per-language. In addition to knowing the names for particular commands, the reversible debugger also needs to be able to query and modify the current state of the debugging session. In particular, the reversible debugger needs a way to determine the current stack depth and the set of active breakpoints. This information is used in the implementation of reverse debugging commands to decided when to convert between `next` and `step` commands.

Each underlying debugger may make this information available in a different format, which needs to be parsed by the overlay debugger. For example, Python's debugger and GDB display stack-frames in opposite orders; a universal debugger must be able to interpret both. Our implementation employs a modular system of debugger "personalities", discussed further in Section 4.

Table 1 describes the basic features of several widely used debuggers. As can be seen, tracing support is ubiquitous and reverse support is rare. Our composite debugging architecture assumes only limited support from each underlying debugger and is, therefore, able to bridge this gap.

### 2.2 Using Replay to Reverse Execution

The reversible debugger needs to interface with the record/replay library but it does not depend on the implementation of that library. In particular, the reversible debugger needs to know how to start recording, and how to rewind and replay from a particular point. Note that record/replay only applies to the debugged program and not to the underlying debugger itself. The underlying debugger does not know about snapshots or reverse execution.

Figure 1 illustrates two cases that are important to distinguish. First, some debuggers, such as GDB or LLDB, are a separate process outside the replay environment (left of Figure 1). Second, for interpreted languages like Python or Perl, the debugger may be part of the interpreter and therefore inside the target process, which is part of the replay environment (right of Figure 1). This distinction is important: if the debugger is inside the replay environment, we parse the process output to distinguish between debugger output and application output.

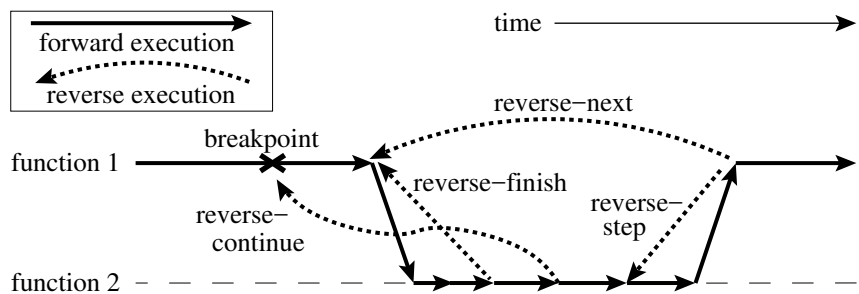

**Figure 2** The various reverse- commands

The reverse commands (Figure 2) `reverse-step`, `reverse-next`, `reverse-finish`, and `reverse-continue` each had to be written with care to avoid subtle algorithmic bugs. The underlying principle is that a `continue` debugging instruction can be expanded into repeated `next` and `step`. Similarly, a `next` can also be expanded into

**Transition Watchpoints**

repeated `step`. Thus, in a typical example, [`continue, next, next, reverse-step`] might expand into [`continue, next, step, next, step, reverse-step`], which then expands into [`continue, next, step, next`]. FReD uses checkpoint-restart to expand one GDB history into another. Algorithm 1 shows the algorithm for reverse-next, which is the most complex case. A previous workshop paper included algorithms for reverse-step and reverse-continue [42].

---

**Algorithm 1:** Reverse-next algorithm. Reverse-finish is depicted in Figure 2; re-execute restarts from the previous checkpoint.)

---

1: **while true do**
2:   **if** *last command is continue or next/bkpt* **then**
3:     *set cmd ← last command*
4:     reexecute without last debugger command
5:     **if** *cmd is next/bkpt and same_stack_depth()* **then**
6:       break
7:     **else if** *cmd is next/bkpt and deeper()* **then**
8:       {`next/bkpt` had exited a function}
9:       execute `reverse-finish`
10:       break
11:     **else**
12:       {else shallower() or cmd is `continue`}
13:       execute `step`
14:       **while** *we are not at breakpoint* **do**
15:         execute `next`
16:       {go to to next iter of while loop}
17:   **else if** *last command is step or next* **then**
18:     reexecute without last debugger command
19:     **if** *same() or shallower()* **then**
20:       break
21:     **else if** *deeper()* **then**
22:       {`next` had exited a function}
23:       execute `reverse-finish`
24:       break

---

## 3 Expression Watchpoints

The previous section described how reversibility can be added to an existing debugger, by using record-replay. Here, we discuss how to add a more sophisticated mechanism, the expression-transition watchpoint. We begin with the basic concept of the algorithm, and then discuss the details of first the single-threaded and then the multi-threaded case.

A conventional watchpoint (in a debugger such as GDB) stops a program when a particular variable is altered. This will execute slowly when implemented in software, or more efficiently when using memory protection hardware. We extend the notion of watchpoint from particular *values* to arbitrary *expressions*. An expression-transition watchpoint stops the execution of a program at the point where a boolean expression changes value. This means that a programmer can jump to the point in execution where an invariant is violated and can then step backwards or examine the program to start to understand exactly what happened.

Our model is that the user can identify a point where the expression is true and a point where the invariant is false, and the user seeks to find the point where the expression changes value. This is motivated by the common case of trying to understand why an invariant is violated during program execution. The invariant is presumably true at some identifiable point, and then is false when the program failed. In the context of a debugging session, the point where the invariant is false is typically the latest point in the execution of the program.

It might happen that the watch expression changes value more than once: that a program invariant is repeatedly violated and then restored, only to be violated again at the latest point in the program execution. An expression-transition watchpoint is guaranteed to identify at least one point where the invariant is broken. We make no claim about which point that will be. A developer seeking to pick a particular transition can add additional clauses to the watch condition.

## 3.1 Overview

A simple implementation of expression-transition watchpoints would check the invariant at each program step. This is very costly if the invariant is not a simple memory value. Without hardware support, value watchpoints can impose a slowdown factor of several hundred. Since page-protection hardware cannot check general invariants, some other implementation technique is required for the sake of efficiency.

Our approach is to exploit checkpoints and determinism to do a binary search over the process history. This avoids the need to check the invariant at each program step. Assume for simplicity that the watch expression is true at the start and false at the end. We can pick an intermediate point in the execution. If the expression is false, then there must have been a transition before this intermediate point; if true, there must be a transition after. By repeatedly bisecting the execution, it is possible to identify the exact program step that caused the expression to change value.

If implemented naively, this binary search will re-execute some statements up to $\log_2 N$ times. We can avoid this by using checkpoints. If the debugger takes a checkpoint at each stage of the binary search, then no particular sub-interval over the time period needs to be executed more than twice.

We do still need logarithmically many evaluations of the watch expression. The actual number of such evaluations will be low, however. Using binary search, the number of expression evaluations will be at most $\log_2 N$, where $N$ is the number of program statements executed. A CPU core running at 2 GHz for five days will execute approximately $10^{15}$ instructions. As a result, it would only take approximately 50 expression evaluations to pin down the precise instruction that caused an expression to change values.

In this way, the typical running time will be bounded by 50 checkpoints, 50 restarts and the time to re-execute the code in the time interval of interest. Checkpoint and restart typically proceed within seconds. So, for a reasonable running time of the code, this implies that the time for a reverse expression watch will be on the order of one minute to 100 minutes. This number is in keeping with the experimentally determined times of Table 4 in Section 5.

**Transition Watchpoints**

### 3.2 The Single-threaded Case

We begin by describing how expression-transition watchpoints work in the single threaded case. If a program is single-threaded, there is a well-defined ordering of program statements and so the algorithm needs only to find the statement that causes the watch expression to change value. We assume that the developer will take an initial checkpoint before the program section of interest. There may be additional programmer-created checkpoints as well. An extension, which we have not implemented, would take additional checkpoints automatically. This is not required for correct behavior and is purely an optimization.

The algorithm for expression-transition watchpoints works in two steps: first, finding a pair of checkpoints that bracket the failure and then finding the culprit statement between the two checkpoints.

(A) *Search-Ckpts:* Binary search over the set of checkpoints to find two successive checkpoint images evaluated as "good" and "bad". It can happen that all previous checkpoint images were "good". In this case, the desired checkpoint interval is from the most recent checkpoint image until the current point in time.

(B) *Search-Debug-History:* A binary search in the debug history between the "good" checkpoint image and the "bad" point in time. As needed in the search through the debugging history, refine "continue" and "next" commands into repeated "step" commands to identify a single "step" command that causes a transition from "good" to "bad". (This relies on the techniques discussed above in Section 2.2)

### 3.3 The Full Algorithm

The multi-threaded case is more complex, since there is now no longer a unique ordering of events. Moreover, existing debugger interfaces often have limited thread support: typically there is one "foreground" thread manipulated by step commands, and additional "background" threads that execute asynchronously. As a result, significant work is required to use the debugger interface to find which statement in which thread is responsible for changing the watch expression. The multi-threaded version of our implementation has been primarily developed with GDB in mind, which offers more complete support for multi-threaded debugging than most other debuggers.

We extend the basic algorithm from Section 3.2 with two additional steps. First, in addition to a binary search over checkpoints and over debugging steps, we also search over the replay event log. Second, we search through the threading schedule to find a deterministically replayable series of debugger commands that allows the end user to observe which thread caused the transition from "good" to "bad".

The binary search over the event log is an optimization to improve efficiency. The deterministic replay library by definition fixes a particular order of events to be replayed; in particular, system calls (such as lock acquisition and release) are all totally ordered. As a result, we can search through this log to bound the period in which the watch expression changed value.

Once a period between two system calls has been isolated by the search over the event log, it remains to pin down a specific statement. The details of this will depend on the way each underlying debugger supports threads. For GDB, we use the debugger's *scheduler-locking* mechanism. Scheduler locking is a mode of operation in which a single "step" command causes just one thread to execute.

When using scheduler locking, background threads do not execute. As a result, there are only two possible outcomes from stepping the program forward. Either some statement execution by the active thread will cause a good-to-bad transition, or else the single step execution will block on a lock, causing deadlock. The deadlock is detected via a timeout of the single step. At this point, the last single step is "undone" (for example, by replaying from the previous checkpoint until the previous step), and another thread is then chosen as the active thread.

(A) *Search-Ckpts:* described above.

(B) *Search-Debug-History:* described above.

(C) *Search-Determ-Event-Log:* Binary search through the portion of the deterministic replay log corresponding to the last "step" command, as identified by Step B. Identify two consecutive events, such that the watched expression transitions from "good" to "bad" when replaying the events. Since multiple threads may have executed, multiple log events may have occurred. Note that a background thread in the target application may be responsible for the transition of the watched expression to "bad". Since the background thread may not yet have been created, a binary search through the event log will guarantee that the execution progresses far enough to guarantee that the background thread exists, since thread creation is one of the events that is logged.

(D) *Local-Search-With-Scheduler-Locking:* Replay the code with scheduler locking. Switch deterministically in a round-robin fashion among the threads of the target application. Execute "step" commands in the active thread. If executing a "step" in the active thread causes the expression to transition from "good" to "bad", this must be the target thread and statement. Else, upon reaching the end of the interval or a deadlock, repeat the process in each other thread.

The actual details of Step D differ for reasons of efficiency:

(D.1) Select a thread as the "active" thread, and do repeated "next" commands to that thread (without scheduler locking) until the expression changes. Then determine if this is the correct thread by re-executing the same series of debugger commands and enabling scheduler locking on the last "next" command and observe if the expression still changes. If it does, we are guaranteed that this is the correct thread. If we see a deadlock, we don't know if this is the right thread. If the expression doesn't change, this is the wrong thread; restart.

(D.2) Undo the last "next", and replace by a single "step" followed by repeated "next" (without scheduler locking). If the expression changes on that first step, go to step D.3 below. If the expression does not change, then restart from D.1 with a different thread.

**Transition Watchpoints**

(D.3) The expression changed on this "step". We must verify that it is due to the active thread. Undo "step", enable scheduler locking, and redo "step." If the expression changes, this is the right thread; exit. If the expression does not change, or if deadlock ensues, this is not the right thread; restart from D.1.

Our current implementation uses a timeout (currently 20 seconds) in order to decide if a deadlock occurred inside step (D.1) or step (D.3). A more sophisticated approach would look at the point where each thread is stopped; if the active thread is blocked waiting for a lock release, this implies that this was the wrong thread schedule to reproduce the problem.

## 4 The FReD Implementation

Section 2 described composite debugging as a design technique. Here, we discuss the FReD implementation in particular, with an eye to the aspects that required significant thought to get right, or that expose the strengths and weaknesses of the architecture as a whole.

The FReD implementation is composed of three layers: the checkpoint/restart system, the record/replay system and the FReD application. For the checkpointing system, FReD uses Distributed MultiThreaded CheckPointing (DMTCP) [2], a transparent user-level checkpointing package. The record/replay system is implemented as a plugin for DMTCP. Finally, the top-level FReD application, which the user interacts with, is implemented in Python. Within the FReD application, each supported debugger has a small amount of custom code called a "personality". The complexity of the record/replay plugin and FReD application in terms of lines of code is represented in Table 2. We now discuss each layer in more detail.

■ **Table 2** Number of lines of code for the major components in FReD, measured using "sloccount."

| Component | # lines of code | Language |
|---|---:|---|
| Ptrace plugin | 935 | C/C++ |
| Record/replay plugin | 8,071 | C/C++/Python |
| FReD application | 3,624 | Python |
| GDB personality | 200 | Python |
| MATLAB personality | 97 | Python |
| Perl personality | 82 | Python |
| Python personality | 72 | Python |

### 4.1 Personalities

FReD incorporates modular debugger "personalities" to cope with the vagaries of different underlying debuggers. The personality module for each debugger is responsible for translating higher-level commands such as "step over," "set breakpoint" or "backtrace" into the debugger-specific versions (such as `next`, `break`, and `where` in the case of GDB). Currently, FReD supports GDB, Python, Perl, and partly supports MATLAB's builtin command-line debugger.

For debuggers supporting different features, the personality is also responsible for translating unsupported commands into a semantically equivalent sequence of supported commands. For example, Python's debugger does not support "count" arguments such as "`next 5`" to perform five sequential steps. The personality translates a "`next 5`" command issued by FReD into a series of five `next` commands.

This modular approach allows great flexibility in supporting new debuggers with FReD. All that is required to support a new debugger is the addition of a new personality with functions to parse the underlying debugger output.

The required engineering is straightforward: Table 2 lists the size of each of the currently supported personalities. The personality for GDB is much larger than the others. This extra code is required to support multithreaded debugging with GDB using scheduler-locking. The other underlying debuggers have limited or no multithreaded debugging support. In our experience, each new personality can be implemented in approximately four hours. Listing 1 shows a snippet of the Python personality.

■ **Listing 1** Code snippet of Python personality. Repetitive definitions as well as pretty-printing helper definition have been left out for compactness. See [3] for a complete listing.

```python
1  class PersonalityPython(personality.Personality):
2    def __init__(self):
3      personality.Personality.__init__(self)
4      # Python equivalent for next, step, etc.
5      self.PROMPT = "(Pdb) "
6      self.NEXT = "next"
7      <similarly for step, continue, where, etc.>
8      ...
9      # Regular expressions for parsing backtrace, etc.
10     self.re_prompt = re.compile("\(Pdb\) $")
11     self.re_backtrace_frame = \
12       ".+/(.+?)\((\d+)\)(.+?)\(.*?\).*?\n–\>"
13     self.re_breakpoint = \
14       "(\d+)\s+(\w+)\s+(\w+)\s+(\w+)\s+" \
15       "at .+/(.+):(\d+)\s+" \
16       "(?:breakpoint already hit (\d+) time)?"
17
18     # Things like 'next 5' are not allowed:
19     self.b_has_count_commands = False
20
21     # Return a BacktraceFrame from the given RE object
22     def _parse_backtrace_frame(self, match_obj):
23       frame = debugger.BacktraceFrame()
24       frame.s_file = match_obj[0]
25       frame.n_line = int(match_obj[1])
26       frame.s_function = match_obj[2]
27       return frame
28
29     # Return a Breakpoint from the given RE object
30     def _parse_one_breakpoint(self, match_obj):
31       breakpoint = debugger.Breakpoint()
32       breakpoint.n_number = int(match_obj[0])
33       breakpoint.s_type = match_obj[1]
34       breakpoint.s_display = match_obj[2]
35       breakpoint.s_enable = match_obj[3]
36       breakpoint.s_file = match_obj[4]
37       breakpoint.n_line = int(match_obj[5])
38       breakpoint.n_count = to_int(match_obj[6])
39       return breakpoint
```

### 4.2 Checkpoint-Restart

DMTCP is used to checkpoint and re-execute the entire debugging session. DMTCP works completely in the user space without requiring any kernel or application modifications. It uses `LD_PRELOAD` to inject the checkpointing library in the target processes. A centralized coordinator process is used to coordinate checkpointing over multiple processes. Further, DMTCP's plugin architecture allows one to provide application-specific customizations to the checkpointing process.

Debuggers for interpreted languages such as Python, Perl, and Matlab run as part of the interpreter, which provides a special mode for tracing and debugging the application program. These can be checkpointed directly, capturing both the state of the target program and of the debugger.

**Transition Watchpoints**

Other debuggers, such as GDB, use the `ptrace` system call to trace the program execution in a separate process. The `ptrace` system call allows a *superior* process (debugger) to inspect and modify the state of an *inferior* process (application).

While DMTCP can do coordinated checkpoint-restart of multiple processes, it cannot directly checkpoint uses of the `ptrace` system call, since we cannot force the inferior to perform any action while it is being traced by a superior.

We overcame this limitation by creating a ptrace plugin for DMTCP that creates a wrapper for the `ptrace` calls. At the time of checkpoint, the plugin forces the superior to detach the inferior process, allowing DMTCP to complete the checkpoint. Once the checkpoint is complete, the plugin forces the superior to once again attach to the inferior.

### 4.3 Deterministic Record-Replay

There are several potential sources of nondeterminism in program execution, and record-replay must address all of them: thread interleaving, external events (I/O, etc.), and memory allocation. While correct replay of external events is required for all kind of programs, memory accuracy is often not an issue for higher-level languages like Python and Perl, which do not expose the underlying heap to the user's program.

FReD handles all these aspects by wrapping various system calls. Relevant events are captured by interposing on library calls using `dlopen/dlsym` for creating function wrappers for interesting library functions. The wrappers record events into the log on the first execution and then return the appropriate values (or block threads as required) on replay.

We start recording when directed by FReD (often after the first checkpoint). The system records the events related to thread-interleaving, external events, and memory allocation into a log. On replay, it ensures that the events are replayed in the same order as they were recorded. The plugin guarantees deterministic replay — even when executing on multiple cores — so long as the program is free of data races.

**Thread interleaving** FReD uses wrappers around library calls such as `pthread_mutex_lock` and `pthread_mutex_unlock`, to enforce the correct thread interleaving during replay. Apart from the usual `pthread_xxx` functions, some other functions that can enforce a certain interleaving are blocking functions like `read`. For example, a thread can signal another thread by writing into the write-end of a pipe when the other thread is doing a blocking read on the read-end of the pipe.

**Replay of external events** Applications typically interact with the outside world as part of their execution. They also interact with the debugger and the user, as part of the debugging process. Composite debugging requires separating these streams. For debuggers that trace a program in a separate process, the I/O by the process being debugged is recorded and replayed whereas the I/O by the debugger process is ignored.

For interpreted languages, the situation becomes trickier as the record-replay plugin cannot differentiate between the debugger I/O and the application I/O. FReD handles this situation heuristically. It designates the standard input/output/error file descriptors as pass-through devices. Activity on the pass-through devices is ignored by the record-replay component.

**Memory accuracy** One important feature of FReD is *memory-accuracy*: the addresses of objects on the heap do not change between original execution and replay. This is important because it means that developers can use address literals in expression watchpoints (assuming they are supported by the underlying debugger).

With true replay of application program, one would expect the memory layout to match the record phase, but the DMTCP libraries have to perform different actions during normal run and on restart. This results in some memory allocation/deallocations originating from DMTCP libraries that can alter the memory layout. Another cause for the change in memory layout is the memory allocated by the operating system kernel when the process doesn't specify a fixed address. An example is the `mmap` system call without any address hint. In this case, the kernel is free to choose any address for the memory region.

Memory-accuracy is accomplished by logging the arguments, as well as the return values of `mmap`, `munmap`, etc. on record. On replay, the real functions or system calls are re-executed in the exact same order. However, the record-replay plugin provides a hint to the kernel to obtain the same memory address as was received at record-time. FReD handles any conflicts caused by memory allocation/deallocation originating from DMTCP itself by forcing use of a separate allocation arena for DMTCP requests.

## 5 Motivating Synthetic Test

Our evaluation focuses on three questions: performance, utility, and flexibility. This section evaluates the performance, as compared against the use of GDB software watchpoints, embedded assert statements in the target program, and expression-transition watchpoints. The benefits of logarithmic scaling for expression-transition watchpoints are experimentally demonstrated. (In fact, GDB software watchpoints turn out to be too slow for even the smallest test cases.) Section 6 evaluates the utility through four case studies concerned with reported bugs in MySQL, Firefox, and Pbzip2.

The performance tests were conducted on a dual-core Intel i7-2640M laptop computer with 8GB RAM and Intel 320 series solid state disk. This system was running OpenSUSE 13.1 with kernel 3.11.6-4-desktop. The experiments described in Section 6 were carried out on a 16-core computer with 128GB of RAM. The computer has four 1.80 GHz Quad-Core AMD Opteron Processor 8346 HE and it runs Ubuntu version 11.10 with kernel 3.0.0-12-generic. The kernel, glibc, GDB and gcc were unmodified.

We begin by showing that FReD performance scales well as application instances grow. A C program was developed that adds edges to a large graph. The imagined application assumes that the graph will always be acyclic. However, one of the edges being inserted creates a cycle. The goal is to identify that edge. (Efficient algorithms for cycle detection exist, but the goal of this exercise is to see how well FReD scales.)

**Transition Watchpoints**

■ **Table 3** The time (in seconds) required by FReD to find the faulty statement that inserted a cycle in a directed acyclic graph. $N$ is the graph size with 34,546 nodes and 421,576 edges from [15]. $2N$, $4N$ and $8N$ represent graphs with a multiple of this number of nodes and edges. Note that the time for reverse expression watch doesn't grow linearly with the increase in graph size. Reverse expression watch was invoked as "`fred-rev-watch has_cycle() == 1`".

| Graph Size | Native Runtime | Runtime w/ FReD | Runtime w/ Assert | Time to complete Rev. Exp. Watch |
|---|---|---|---|---|
| $N$  | 0.010 | 0.012 | 98.458   | 15.125 |
| $2N$ | 0.018 | 0.020 | 228.094  | 15.159 |
| $4N$ | 0.032 | 0.041 | 525.594  | 18.748 |
| $8N$ | 0.066 | 0.085 | 1058.033 | 19.143 |

Three debugging options are available. One option is to run the program under GDB using its software watchpoint facility to check for a cycle upon executing each statement. A hardware watchpoint cannot be used here, and *software watchpoints are too slow to finish in less than a day*. So this first option is not be represented in Table 3.

A second option is to add an assert statement into the program to check for the existence of a cycle after each edge is added. The third option is to use FReD's expression-transition watchpoint to find the point in program execution when a cycle was added. In both cases, a small, straightforward function that does a depth-first search in testing for the existence of a cycle is used. This is invoked either in the assert statement or by FReD's expression-transition watchpoint facility.

The results for the latter two debugging strategies are summarized in Table 3. As shown, the runtime with assert statements grows approximately linearly with the problem size. However, due to binary search, the runtime with expression-transition watchpoints grows logarithmically with the problem size.

## 6 Case Studies

In the previous section, we showed that our implementation is efficient. Here, we show that it is useful, by showing how it would help debug various real problems. We picked these bugs by looking for complex failures in multi-threaded programs. Each bug demonstrates a different real-world scenario.

Due to our unfamiliarity with the code bases for these programs, we chose bugs for which patches had already been submitted. This allowed us to more easily understand the circumstances of the bug, and in some cases led to an expression suitable for a transition watchpoint.

The case studies presented here use C/C++ programming language. The debugger being used was GDB. For each of the following MySQL examples, the average number of entries in the deterministic replay log was approximately 1.2 million. The average size of an entry in the log was approximately four bytes.

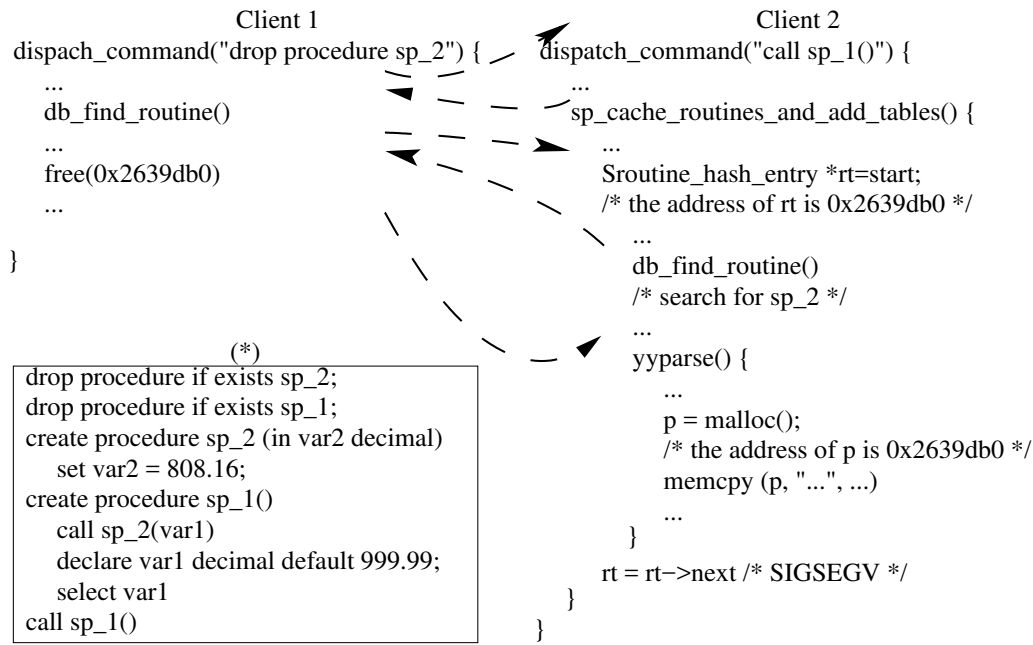

**Figure 3** MySQL Bug 12228: the thread interleaving that causes the MySQL daemon to crash with SIGSEGV; (*) the sequence of instructions executed by each thread, in pseudo-SQL

## 6.1 MySQL Bug 12228 — Atomicity Violation

MySQL bug 12228 was a report that a particular multi-threaded stress test was failing. In this test, ten threads issued concurrent client requests to the MySQL daemon. After a period of execution on the order of tens of seconds, the MySQL daemon crashed with a bad pointer dereference. In our experience, this bug occurred approximately 1 time in 1000 client connections. This bug was reproduced using MySQL version 5.0.10.

This bug was a perfect candidate to illustrate the utility of a transition watchpoint. The bug manifestation was the MySQL daemon crashing with a bad pointer dereference. While the bug itself was easy to reproduce, it was not immediately clear how, when or why the pointer value had been set to its bad value. However, by using a transition watchpoint, we were immediately able to locate the exact statement and thread interleaving that resulted in the pointer being set to the value resulting in a crash.

The buggy thread interleaving and the series of requests issued by each client are presented in Figure 3. The bug occurs when one client, "client 1" removes the stored procedure sp_2(), while a second client, "client 2" is executing it. The memory used by procedure sp_2() is freed when client 1 removes it. While client 1 removes the procedure, client 2 attempts to access a memory region associated with the now non-existent procedure. Client 2 is now operating on unclaimed memory. The MySQL daemon is sent a SIGSEGV.

This bug was diagnosed with FReD in the following way: the user runs the MySQL daemon under FReD and executes the stress test scenario presented in Figure 3. The debug session is presented below. Some debugger output has been omitted for clarity.

**Transition Watchpoints**

```
(gdb) continue
Program received signal SIGSEGV.
in sp_cache_routines_and_table_aux at sp.cc:1340
sp_name name(rt->key.str, rt->key.length)
(gdb) print rt
$1 = 0x1e214a0
(gdb) print *rt
$2 = 1702125600
(gdb) fred-reverse-watch *(0x1e214a0) == 1702125600
FReD: 'fred-reverse-watch' took 406.24 seconds.
(gdb) list
344     memcpy(pos,str,len);
```

When the SIGSEGV is hit, GDB prints the file and line number that triggered the SIGSEGV. The user prints the address and value of the variable rt. The value of rt is "bad", since dereferencing it triggered the SIGSEGV. From there it is a simple conceptual problem: at what point did the value of this variable rt change to the "bad" value? FReD's reverse expression watchpoint (or fred-reverse-watch as abbreviated above) is used to answer this question. In the case of this bug, an unchecked memcpy() call was overwriting the region of memory containing the rt pointer, leading to the SIGSEGV.

The time for reverse expression watchpoint, as well as other useful information, are shown in Table 4.

■ **Table 4** The bugs and the time it took FReD to diagnose them, by performing a expression-transition watchpoint (in seconds). Other timings that are of interest are shown: the total and average times for checkpoint, restart and evaluation of the expression (in seconds), as well as the number of checkpoints, restarts and evaluation of the expression.

| Bug | Native Runtime (s) | Rev-Watch (s) | Ckpt (#) | Rstr (#) | Ckpt (s) | Rstr (s) | Eval Expr (s) | Eval Expr (#) |
|---|---|---|---|---|---|---|---|---|
| MySQL 12228 | 11.49 | 406.24 | 4 | 60 | 3.45 | 24.49 | 1.69 | 93 |
| MySQL 42419 | 27.37 | 161.68 | 6 | 55 | 6.17 | 22.59 | 1.06 | 91 |
| pbzip2 | 1.31 | 29.22 | 1 | 17 | 0.99 | 5.60 | 0.41 | 27 |

**6.2 MySQL Bug 42419 — Data Race**

In this bug, the MySQL daemon crashed with a SIGABRT caused by attempting to jump to an invalid location. The cause of the bug was attempting to execute a function eq() whose address was invalid. However, it was not immediately obvious that the address was invalid, and thus some additional insight was needed. A transition watchpoint in this scenario was the perfect tool to lead us to just before the bug occurred to allow us to understand the exact sequence of events causing the crash.

In order to reproduce MySQL bug 42419, two client threads that issue requests to the MySQL daemon (version 5.0.67) were used, as indicated in the bug report. MySQL bug 42419 was diagnosed with FReD. The debug session is shown next:

```
(gdb) continue
Program received signal SIGABRT at sql_select.cc:11958.
if (ref_item && ref_item->eq(right_item, 1))
(gdb) print ref_item
$1 = 0x24b9750
(gdb) print table->reginfo.join_tab->ref.items[part]
$2 =  0x24b9750
(gdb) print &table->reginfo.join_tab->ref.items[part]
$3 = (class Item **) 0x24db518
(gdb) fred-reverse-watch *0x24db518 == 0x24b9750
```

The crash (receiving a SIGABRT) was caused by the fact that the object `ref_item` did not contain a definition of the `eq()` function. We began by using `reverse-watch` to identify the statement where `ref_item` was last modified, to hopefully understand how `eq()` became invalid. The modification of `ref_item` happens during a call to the function `make_join_statistics():sql_select.cc:5295` at instruction `j->ref.items[i]=keyuse->val`.

We then step through `make_join_statistics()` using the standard GDB `next` command, and watch MySQL handle an error induced by the client threads. In the process handling the error, the thread frees the memory pointed to by `&ref_item`. But, crucially, it does not remove it from `j->ref.items[]`. Thus, when a subsequent thread processes these items, it sees the old entry, and attempts to dereference a pointer to a memory region that has previously been freed. The time for reverse expression watchpoint, as well as other useful information, are shown in Table 4.

### 6.3 Firefox Bug 653672

To illustrate another use case of FReD beyond a transition watchpoint, we demonstrate using the reverse step ability of FReD to recover a usable stack trace after the stack had been corrupted. This was a bug in the Firefox Javascript engine (Firefox version 4.0.1). The bug was reproduced using the test program provided with the bug report. The Javascript engine was not correctly parsing the regular expression provided in the test program and would cause a segmentation fault. The code causing the segmentation fault was just-in-time compiled code and so GDB could not resolve the symbols on the call stack, causing an unusable stack trace.

```
(gdb) continue
Program received signal SIGSEGV, Segmentation fault.
(gdb) where
#0  0x00007fffdbaf606b in ?? ()
#1  0x0000000000000000 in ?? ()
(gdb) fred-reverse-step
FReD: 'fred-reverse-step' took 6.881 seconds.
(gdb) where
#0  JSC::Yarr::RegexCodeBlock::execute (...) at yarr/yarr/RegexJIT.h:78
#1  0x7ffff60e3fbb in JSC::Yarr::executeRegex (...) at yarr/...
#2  0x7ffff60e47b3 in js::RegExp::executeInternal (...) at ...
...
```

**Transition Watchpoints**

Here, we used FReD's "reverse-step" to return to the last statement for which the stack trace was still valid. The "reverse-step" took 6.9 seconds. Having a usable stack trace is often crucial information to aid in debugging. In this case, "reverse-step" was a natural tool to use: as soon as the stack is corrupted, simply step back once to the statement causing the corruption. Alternatively, if the stack corruption had occurred long before the SIGSEGV, then a transition watchpoint on whether the stack was valid would have been the preferred approach.

### 6.4 Pbzip2 — Order Violation

Finally, we illustrate another use case of a transition watchpoint. This was a simple case of an auxiliary thread freeing a global variable before another thread attempts to use it. A transition watchpoint delivered us to the exact point in time where the variable was freed. Without a transition watchpoint, a tedious repetitive debugging technique would be needed in order to identify why the variable was being freed before its use.

pbzip2 decompresses an archive by spawning consumer threads which perform the decompression. Another thread (the output thread) is spawned, which writes the decompressed data to a file. Only the output thread is joined by the main thread. Therefore, it might happen that when the main thread tries to free the resources, some of the consumer threads have not exited yet. A segmentation fault is received in this case, caused by a consumer thread attempting to dereference the NULL pointer. The time for reverse watch is shown in Table 4. The debugging session is presented below:

```
(gdb) continue
Program received signal SIGSEGV at
pthread_mutex_unlock.c:290.
(gdb) backtrace
#4 consumer (q=0x60cfb0) at pbzip2.cpp:898
...
(gdb) frame 4
(gdb) print fifo->mut
$1 = (pthread_mutex_t *) 0x0
(gdb) p &fifo->mut
$2 = (pthread_mutex_t **) 0x60cfe0
(gdb) fred-reverse-watch *0x60cfe0 == 0
```

## 7  Discussion

This paper has shown how to build reversible debuggers, with transition watchpoints, on top of an existing debugger plus a deterministic replay framework. This section discusses the intrinsic limits to our approach, and also the aspects that we believe can be improved by future work.

FReD assumes that the threads of the application in question do not access shared memory unless the access is protected by a lock. The call to lock-related system calls is

■ **Table 5** Among checkpoint/re-execute based reversible debuggers, other examples are limited to examining single addresses, and do not support general expressions.

| Reversible Debugger | Multi Threaded | Multi Core | Rev. Exp. Watchpoint | Observations |
| --- | --- | --- | --- | --- |
| IGOR [13] | No | No | x > 0 | only monotonously varying single variables |
| Boothe [7] | No | No | x > 0 | only probes where the debugger stops |
| King et al. [18] | Yes | No | x | detects the last time a variable was modified |
| FReD | Yes | Yes | Complex expressions | detects the exact instruction that invalidates the expression |

then logged, guaranteeing deterministic replay. Some code may omit the lock around shared access (either as a bug, or else on purpose in cases where programmers feel that they can write more efficient code by ignoring these best practices).

Some CPU instructions are troublesome for replay technology. For example, the Intel/AMD rdtsc instruction (read time stamp counter) may be used instead of the gettimeofday system call. In such cases, the application binary will have to be modified to capture the results of these instructions for deterministic replay. Fortunately, replay technology is a steadily-advancing field and we expect that technical support for determinism will keep pace with these hardware developments.

Nevertheless, note that the general approach continues to work well for most compiler optimizations such as just-in-time compilers and instruction caches. As long as a debugging interface for the language in question is provided, and as long as that interface supports the concepts of stopping at a statement or function/procedure, and displaying the current program stack, a FReD personality can be developed for that language.

The performance of our implementation can be improved in a number of ways. The current replay logs are more voluminous than necessary. The average size of a log entry is 79 bytes for the MySQL testbed. 90% of those entries are for pthread_mutex_lock/unlock. A compact representation of that common case would reduce the size to 8 bytes or less. Additionally, each log entry includes extra fields used for debugging. The general entry would be reduced to 20 bytes or less by adding a non-debugging mode.

The approach used in detecting deadlocks can also be improved. Currently, FReD detects deadlocks heuristically, by waiting to see if the executing GDB command completes in 20 seconds. While this has worked fine in practice, it would be possible to have a precise and sound approach by analyzing the waits-for graph of the program.

## 8 Related Work

In this section, we compare FReD with other systems that implement reverse watchpoints (for single variables, rather than expressions; see Section 8.1) or other reversible debuggers (Section 8.2). Deterministic replay systems are also discussed. (Section 8.3).

**Transition Watchpoints**

### 8.1 Reverse Expression Watchpoints

Table 5 presents other reversible debuggers that support even a limited form of transition watchpoint. This includes debuggers that support only a single variable (such as a single hardware address).

Both IGOR [13] and the work by Boothe [7] support a primitive type of reverse expression watchpoint for single-threaded applications of the form x>0, where the left-hand side of the expression is a variable and the right-hand side is a constant. In that work, x must also be a monotonously increasing or decreasing variable. In contrast, our work supports general expressions.

In terms of how a transition watchpoint is performed, IGOR locates the last checkpoint before the desired point and re-executes from there. Boothe performs a transition watchpoint in two steps: first recording the last "step point" at which the expression is satisfied and then re-executing until that point. A step point is a point at which a user-issued command stops. In other words, Boothe can only probe the points where the debugger stops. But a `continue` command can execute many statements. FReD, on the other hand, brings the user directly to a statement (one that is not a function call) at which the expression is correct, for which executing the statement will cause the expression to become incorrect.

The work of King et al. [18] goes back to the last time a variable was modified, by employing virtual machine snapshots and event logging. While the work of King et al. detects the last time a variable was modified, FReD takes the user back in time to the last point an expression had a correct value. Similarly to Boothe [7], the transition watchpoint is performed in two steps and only the points where the debugger stops are probed. UndoDB [36] implements a similar approach to FReD, but only for one debugger (a modified version of GDB) and uses in-memory live checkpoints.

Whyline [19] allows the programmer to ask "why" or "why not" questions about program execution, but does not allow specification of general expressions. The authors also state it is not designed for debugging program executions exceeding several minutes.

### 8.2 Reversible Debuggers

Four different approaches to building a reversible debugger have been described in the past [41]:

- *record/reverse-execute*: for each executed instruction, the record phase logs the change in state of registers, memory, I/O, etc. On reverse-execute, the logged information is used to *undo* the effects of the given instruction by restoring the state as it existed prior to executing the instruction.
- *record/replay*: uses virtual machine snapshots and event logging. A reproducible clock is achieved by tracking certain CPU counters, such as the number of loads and store since startup. This allows asynchronous events to be replayed according to the time of the original clock.

■ **Table 6** The four primary approaches to reversible debugging. Low, Medium, and High are defined in Section 8.2. A debugger is labeled as *backwards compatible* if it requires no modifications to the kernel, compiler or interpreter.

| Reversible debugger | Info captured | Multi thread | Multi-core on replay | Fwd exec. speed | Rev exec. speed | Bkwd compat |
|---|---|---|---|---|---|---|
| Record/Reverse-Execute Debuggers | | | | | | |
| AIDS [16] | | No | No | | | No |
| Zelkowitz [47] | | No | No | | Depends | No |
| Tolmach et al. [39] | High | No | No | Slow | on | No |
| GDB [14] | | Yes | No | | Cmd | Yes |
| TotalView '11 [40] | | Yes | Yes | | | Yes |
| Record-Replay Debuggers | | | | | | |
| King et al. [18] | Low | Yes | No | Fast | Slow | No |
| Lewis et al. [25] | Low | Yes | No | Fast | Slow | No |
| Post-mortem Debuggers | | | | | | |
| Omniscient dbg [8, 26, 33] | Medium | Yes | (*) | Slow | (*) | No |
| Tralfamadore [23] | Medium | Yes | (*) | Medium | (*) | No |
| Checkpoint/Re-Execute Debuggers | | | | | | |
| UndoDB [36] | Medium | Yes | No | Medium | Medium | No |
| IGOR [13] | | No | No | | | No |
| Boothe [7] | | No | No | | | No |
| Flashback [38] | Medium | No | No | Medium | Medium | No |
| ocamldebug [24] | | No | No | | | No |
| FReD | Medium | Yes | Yes | Medium | Fast | Yes |

- *checkpoint/re-execute*: typically uses *live* checkpoints by forking off child processes of the debugged application at interesting *events*. Events are interesting locations in the code that are padded with special logging instructions, using source transformation.
- *post-mortem debugging*: uses an on-disk database to log all events of interest until the application is terminated. Debugging is then performed only using the event database and without requiring a process.

Table 6 groups FReD and previous reversible debuggers according to the approach taken to build a reversible debugger. Each different approach can be characterized along several dimensions: the amount of information captured during forward execution (Table 6, column 3), whether it supports multithreaded target applications (column 4), whether multithreaded applications can make use of multiple cores for performance on replay (column 5), the forward execution speed (column 6), the reverse execution speed (column 7) and backwards compatibility (column 8).

We classified the amount of information captured during forward execution as: Low (determinism enforced without data capture), Medium (enough information is stored to guarantee deterministic replay) or High (logging the state after each instruction is executed). Similarly, we labeled forward execution speeds as Slow (due to excessive logging), Medium (for reversible debuggers that capture enough information to guarantee deterministic replay) and Fast (native speed via the use of virtual machines). Finally, reverse execution speeds can be Slow (due to large

**Transition Watchpoints**

memory footprints), Medium (due to the deterministic replay strategy), Fast (via checkpoints and binary search) or can depend on the type of reverse command issued (reverse-continue and reverse-next tend to be slow, while reverse-step is fast).

FReD is most similar to "post-mortem" debuggers, which allow a user to examine an execution trace after a process is terminated. For these debuggers, the reverse execution speed is meaningless, since the process no longer exists. FReD operates on a live process, meaning in particular that the user can start debugging before the process has run to completion or seen all possible inputs. Further, our approach is language-agnostic, whereas post-mortem debuggers typically require a specialized debugger.

## 8.3 Deterministic Replay

Deterministic replay is a prerequisite for any reversible debugger that wants to support multithreaded applications. There are many systems that implement deterministic replay in the literature, through a variety of mechanisms: [1, 6, 9, 11, 12, 13, 17, 20, 21, 22, 25, 28, 29, 31, 32, 35, 37, 38, 43, 44, 45, 46]. There are also many systems whose goal is to make the initial execution deterministic [5, 10, 27, 30, 34]. It may be possible to employ one of these systems in the future, but at present, they are not sufficiently integrated with the use of standard debuggers such as GDB. Moreover, the deterministic-replay layer needs to be integrated with the checkpoint library, so that it is feasible to resume execution from a point other than the beginning of the process.

## 9 Conclusion

We have demonstrated a fast and easy-to-write reversible debugger built using snapshots with record-replay of events that cause non-determinism. The reversible debugger is built on top of existing conventional debuggers with specific small personalities. The personalities allow the reversible debugger to recognize program-specific statements, functions/procedures, and stacks of procedures, in order to roll back and roll forward within the given language.

An an expression-transition watchpoint using binary-search over process lifetime is built on top of that infrastructure. Our implementation, FReD, is robust enough to support reversible debugging in such complex, and highly multithreaded, real-world programs as MySQL and Firefox.

**Acknowledgements**  This work was partially supported by the National Science Foundation under Grant ACI-1440788.

**Transition Watchpoints**

## About the authors

**Kapil Arya** is a Distributed Systems Engineer at Mesosphere, Inc. He is an Apache Mesos committer and member of the PMC. He currently holds a Visiting Computer Scientist position at the College of Computer and Information Science at Northeastern University. His research interests include distributed systems, high-performance computing, fault-tolerance, and debugging. He received his Ph.D. from Northeastern University and his B.Sc. from Jai Narain Vyas University. He can be reached at kapil@mesosphere.io.

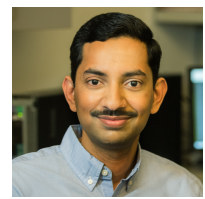

**Tyler Denniston** received his B.S. from Northeastern University in 2012 and his S.M. from MIT in 2016. His areas of research interest include compilers, programming languages, debuggers, distributed and parallel systems, performance optimization, operating systems, virtual machines, and checkpointing. He can be reached at tyler@csail.mit.edu.

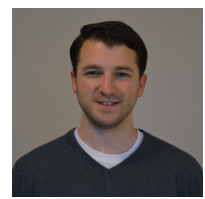

**Ariel Rabkin** is a researcher and engineer primarily interested in monitoring and debugging. He is currently a member of the Support Tools Team at Cloudera, inc, where he works to understand and automate the support process. He received his Ph.D. from UC Berkeley in 2012, and then spent two years as a visiting researcher at Princeton University. He can be reached at asrabkin@gmail.com.

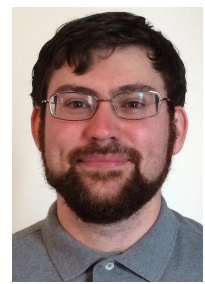

**Gene Cooperman** is a Professor of Computer Science in the College of Computer and Information Science at Northeastern University. He is currently an IDEX Chair of Attractivity at Université Fédérale Toulouse Midi-Pyrénées, France. Prof. Cooperman also has a 15-year relationship with CERN, where his work on semi-automatic thread parallelization of task-oriented software is included in the million-line Geant4 high-energy physics simulator. He research interests include high-performance computing, cloud computing, engineering desktops (license servers, etc.), GPU-accelerated graphics, GPGPU computing, and Internet of Things. He leads the DMTCP project (Distributed MultiThreaded CheckPointing) which includes studying the limits of transparent checkpoint-restart. He received his B.S. from U. of Michigan, and his Ph.D. from Brown University. He can be reached at gene@ccs.neu.edu.

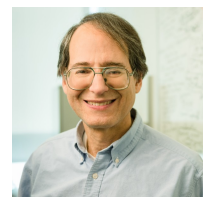